\documentclass[a4paper,twoside]{article}      
\usepackage{amsmath,amssymb,amsfonts,amsthm,amscd} 
\usepackage{graphics}                 
\usepackage{color}                    
\usepackage{ mathrsfs }
\usepackage{indentfirst}
\usepackage{bbold}
\usepackage{enumerate}
\usepackage{url}         
\usepackage{colonequals} 
\usepackage{a4wide}

\newtheorem{theorem}{Theorem}[section]

\theoremstyle{definition}
\newtheorem{remark}[theorem]{Remark}

\def\!{\mathop{\mathrm{!}}}

\newcommand{\Keywords}[1]{\par\indent
{\small{\textbf{Key words.} \/} #1}}
\newcommand{\subjclass}[1]{\par\indent{ \textbf{PACS.} #1}}

\def\R{\mathbb{ R}}

\def\w{\textbf{w}}
\def\z{\textbf{z}}

\def\aL{\mathsf L}
\def\aM{\mathsf M}
\def\aE{\mathsf E}
\def\aS{\mathsf S}
\def\aZ{\mathsf Z}
\def\az{\mathsf z}
\def\aF{\mathsf F}
\def\aG{\mathsf G}

\newlength{\boxwidth}
\date \today
\title{Non-equilibrium thermodynamics of damped Timoshenko and damped Bresse systems}
\author{ Manh Hong Duong
\vspace*{2mm}
\\
Mathematics Institute,\\
University of Warwick, \\
Coventry CV4 7AL, UK. \\
Email: m.h.duong@warwick.ac.uk}
\begin{document}
\maketitle
\begin{abstract}
In this paper, we cast damped Timoshenko and damped Bresse systems into a general framework for non-equilibrium thermodynamics, namely the GENERIC (General Equation for Non-Equilibrium Reversible-Irreversible Coupling) framework. The  main ingredients of GENERIC consist of five building blocks: a state space, a Poisson operator, a dissipative operator, an energy functional, and an entropy functional. The GENERIC formulation of damped Timoshenko and damped Bresse systems brings several benefits. First, it provides alternative ways to derive thermodynamically consistent models of these systems by constructing building blocks instead of invoking conservation laws and constitutive relations. Second, it reveals clear physical and geometrical structures of these systems, e.g., the role of the energy and the entropy as the driving forces for the reversible and irreversible dynamics respectively. Third, it allows us to introduce a new GENERIC model for damped Timoshenko systems that is not existing in the literature.
\end{abstract}
\section{Introduction}
GENERIC (General Equation for Non-Equilibrium Reversible-Irreversible Coupling~\cite{Oettinger05}) is a formalism for non-equilibrium thermodynamics which unifies both reversible and irreversible dynamics. The  main ingredients of GENERIC consist of five building blocks: a state space $\aZ$, a Poisson operator $\aL$, a dissipative operator $\aM$, an energy functional $\aE$, and an entropy functional $\aS$, which are required to satisfy certain conditions. A consequence of these conditions is that the first and the second law of thermodynamics are fulfilled, i.e., the total energy is preserved and the entropy is increasing in time, see Section~\ref{sec: GENERIC}. To show a particular realization of GENERIC one needs to specify the five building blocks and verify the conditions imposed on them. 

GENERIC has been proven to be a powerful framework for mathematical modelling of complex systems. In the original papers~\cite{OG97,OG97part2}, it was originally introduced in the context of complex fluids with applications to the classical hydrodynamics and to non-isothermal kinetic theory of polymeric fluids. Since then many models have been shown to have a GENERIC structure. Recently it has been applied further to anisotropic inelastic solids~\cite{HutterTervoort08a}, to viscoplastic solids~\cite{HutterTervoort08b}, to thermoelastic dissipative materials~\cite{Mielke11}, to the soft glassy rheology model~\cite{FI13}, and to turbulence~\cite{Ottinger14}, just to name a few. More recently, the mathematically rigorous study of GENERIC has received a lot of attention. In~\cite{ADPZ11,DLR13,DPZ13b, MPR14, D14b} the authors show that there is a deep connection between GENERIC structures of many partial differential equations including the diffusion equation, the Fokker-Planck equation and the Vlasov-Fokker-Planck equation and the large deviation principle of underlying stochastic processes. The connection provides microscopic interpretation for the GENERIC structures of these equations~\cite{ADPZ11,DLR13,DPZ13b,MPR14,D14b}, gives ideas to construct approximation schemes for a generalized Kramers equation~\cite{DPZ13a} and offers techniques to handle singular limits of partial differential equations~\cite{AMPSV12,D14b,DPS13}. We refer to the original papers~\cite{OG97, OG97part2} and the book~\cite{Oettinger05} for an exposition of GENERIC and the mentioned papers as well as references therein for further information.

The aim of this paper is to show that damped Timoshenko and damped Bresse systems can be cast into the GENERIC framework. Both Timoshenko and Bresse systems are described by wave equations. In damped Timoshenko and damped Bresse systems, irreversible behaviour is introduced via additional, frictional or heat conduction, damping mechanisms. This can be accomplished in various ways.  The two systems play important roles in theory of elasticity and strength of materials~\cite{T53}. They have been studied extensively in the literature from various perspectives such as existence and uniqueness of solutions, their exponential stability and rate of decay, see for instance~\cite{RR02,RFSC05,HK13,SR09,RR08,FRM14} for damped Timoshenko systems and \cite{SA10,BRJ11,FM12,AFSM14,FR10,SB14,LR09} for damped Bresse systems. 

In this paper we focus on the GENERIC structures of the two systems. The formulation in the GENERIC framework has three benefits. Firstly, it provides alternative ways to derive thermodynamically consistent models of these systems by constructing the building blocks $\{\aZ,\aL,\aM,\aE,\aS\}$ instead of invoking conservation laws and constitutive relations. We demonstrate that all previous known models can be constructed in this way using a common procedure. In addition, we also illustrate how to use GENERIC to come up with a new model for the damped Timoshenko system. Secondly, thanks to its splitting formulation, the GENERIC framework provides clear physical and geometrical structures for these systems, e.g., the role of the energy and the entropy as the driving forces for the reversible and irreversible dynamics respectively. Thirdly, this has the possibility of application of the recent developments for GENERIC as mentioned in the second paragraph to mathematically analyse the two systems.

%
%
%

The organization of the paper is as follows. In Section~\ref{sec: GENERIC}, we review the GENERIC framework and its basic properties. In Section \ref{sec: modelling}, we introduce a procedure for mathematically modelling of complex systems using GENERIC. In Section~\ref{sec: Timoshenko}, we place damped Timoshenko systems with damping either by frictional mechanisms or by heat conduction of various types in the GENERIC framework. A similar analysis for damped Bresse systems is shown in Section~\ref{sec: Bresse}.  Conclusion and further discussion are given in Section \ref{sec: summary}. Finally, Appendix A contains some detailed computations, and Appendix B shows more models of the two systems.
\section{GENERIC}
\label{sec: GENERIC}
In this section, we summarize the GENERIC framework and its main properties. We refer to the original papers~\cite{OG97, OG97part2} and the book~\cite{Oettinger05} for an exposition of GENERIC.
\subsection{Definition of GENERIC}
As mentioned in the introduction, GENERIC is a formulation for non-equilibrium thermodynamics that couples both reversible and irreversible dynamics in a certain way. Let $\az\in \aZ$ be a set of variables which appropriately describe the system under consideration, and $\aZ$ denotes a state space. Let $\aE, \aS\colon \aZ\rightarrow\R$ be two functionals, which are interpreted respectively as the total energy and the entropy. Finally, suppose that for each $\az\in\aZ$, $\aL(\az)$ and $\aM(\az)$ are two operators, which are called Poisson operator and dissipative operator respectively, that map the cotangent space at $
\az$ to  onto the tangent space at $\az$. A GENERIC equation for $\az$ is then given by the following differential equation
\begin{equation}
\label{eq:GENERICeqn1}
\az_t=\aL(\az)\frac{\delta\aE(\az)}{\delta\az}+\aM(\az)\frac{\delta\aS(\az)}{\delta\az},
\end{equation}
where
\begin{itemize}
\item $\az_t$ is the time derivative of $\az$.
\item $\frac{\delta\aE}{\delta\az}, \frac{\delta\aS}{\delta\az}$ are appropriate derivatives, such as either the Fr\'echet derivative or a gradient with respect to some inner product, of $\aE$ and $\aS$ respectively.
\item $\aL= \aL(\az)$ is for each $\az$ an antisymmetric operator satisfying the Jacobi identity, i.e., for all functionals $\aF,\aG,\aF_i\colon\aZ\rightarrow\R,\ i=1,2,3$
\begin{equation}
\label{def: antisymmetry}
\{\aF,\aG\}_{\aL}=-\{\aG,\aF\}_{\aL},
\end{equation}
\begin{equation}
\label{def:Jacobi}
\{\{\aF_1,\aF_2\}_{\aL},\aF_3\}_{\aL}+\{\{\aF_2,\aF_3\}_{\aL},\aF_1\}_{\aL}+\{\{\aF_3,\aF_1\}_{\aL},\aF_2\}_{\aL}=0,
\end{equation}
where the Poisson bracket $\{\cdot,\cdot\}_{\aL}$ is defined via
\begin{equation}
\label{def:Poisson bracket}
\{\aF,\aG\}_{\aL}\colonequals \frac{\delta\aF(\az)}{\delta\az}\cdot \aL(\az)\,\frac{\delta\aG(\az)}{\delta\az}.
\end{equation}
\item Similarly, $\aM=\aM(\az)$ is for each $\az$ a symmetric and positive semidefinite operator, i.e., for all functionals $\aF,\aG\colon\aZ\rightarrow\R,$
\begin{equation}
\label{def: symmetry}
[\aF,\aG]_{\aM}=[\aG,\aF]_{\aM},
\end{equation}
\begin{equation}
\label{def: pos def}
[\aF,\aF]_{\aM}\geq 0,
\end{equation}
where the dissipative bracket $[\cdot,\cdot]_{\aM}$ is defined by
\begin{equation}
\label{def:dissipative bracket}
[\aF,\aG]_{\aM}\colonequals \frac{\delta\aF(\az)}{\delta\az}\cdot \aM(\az)\,\frac{\delta\aG(\az)}{\delta\az}.
\end{equation}
\item Moreover, $\{\aL,\aM,\aE,\aS\}$ are required to fulfill the degeneracy conditions: for all $\az\in \aZ$,
\begin{equation}
\label{def:degeneracy}
\aL(\az)\,\frac{\delta\aS(\az)}{\delta\az}=0,\quad
\aM(\az)\,\frac{\delta\aE(\az)}{\delta\az}=0.
\end{equation}
\end{itemize}
$\{\aZ,\aL,\aM,\aE,\aS\}$ are called the building blocks and a GENERIC system is then fully characterised by its building blocks.
\subsection{Basic properties of GENERIC}
We recall two basis properties of GENERIC. The first one is that GENERIC automatically verifies the first and second laws of thermodynamics. More precisely, the degeneracy conditions~\eqref{def:degeneracy} together with the symmetries of the Poisson and dissipative brackets~\eqref{def: antisymmetry}-\eqref{def: pos def} ensure that the energy is conserved along a solution
\begin{equation*}
\frac{d\aE(\az(t))}{dt}=\frac{\delta\aE(\az)}{\delta\az}\cdot \frac{d\az}{dt}\overset{\eqref{eq:GENERICeqn1}}{=}\frac{\delta\aE(\az)}{\delta\az}\cdot\left(\aL(\az)\frac{\delta\aE(\az)}{\delta\az}+\aM(\az)\frac{\delta\aS(\az)}{\delta\az}\right)\overset{\eqref{def: symmetry},\eqref{def:degeneracy}}{=}\frac{\delta\aE(\az)}{\delta\az}\cdot\aL(\az)\frac{\delta\aE(\az)}{\delta\az}\overset{\eqref{def:Poisson bracket}}{=}\{\aE,\aE\}_{\aL}\overset{\eqref{def: antisymmetry}}{=}0,
\end{equation*}
and that the entropy is a non-decreasing function of time
\begin{equation*}
\frac{d\aS(\az(t))}{dt}=\frac{\delta\aS(\az)}{\delta\az}\cdot \frac{d\az}{dt}\overset{\eqref{eq:GENERICeqn1}}{=}\frac{\delta\aS(\az)}{\delta\az}\cdot\left(\aL(\az)\frac{\delta\aE(\az)}{\delta\az}+\aM(\az)\frac{\delta\aS(\az)}{\delta\az}\right)\overset{\eqref{def: antisymmetry},\eqref{def:degeneracy}}{=}\frac{\delta\aS(\az)}{\delta\az}\cdot\aM(\az)\frac{\delta\aS(\az)}{\delta\az}\overset{\eqref{def:dissipative bracket}}=[\aS,\aS]_\aM\overset{\eqref{def: pos def}}{\geq} 0.
\end{equation*}

The second property of GENERIC, which is useful when constructing the building blocks, is that it is invariant under coordinate transformations~\cite{OG97}. Let $\az\mapsto \overline{\az}$ be a one-to-one coordinate transformation. The new building blocks $\{\overline{\aL},\overline{\aM},\overline{\aE},\overline{\aS}\} $ are obtained via
\begin{equation}
\label{eq: transform}
\overline{\aE}(\overline{\az})=\aE(\az),\quad \overline{\aS}(\overline{\az})=\aS(\az),\quad \overline{\aL}(\overline{\az})=\left[\frac{\partial (\az)}{\partial (\overline{\az})}\right]^{-1} \aL(\az) \left[\frac{\partial (\az)}{\partial (\overline{\az})}\right]^{-T},\quad \overline{\aM}(\overline{\az})=\left[\frac{\partial (\az)}{\partial (\overline{\az})}\right]^{-1} \aM(\az) \left[\frac{\partial (\az)}{\partial (\overline{\az})}\right]^{-T},
\end{equation}
where $\frac{\partial (\az)}{\partial (\overline{\az})}$ is the transformation matrix, $[\cdot]^{-1}$ is the inverse of $[\cdot]$ and $[\cdot]^{-T}$ denotes the transpose of $[\cdot]^{-1}$. Then the transformed GENERIC system $\{\overline{\aZ},\overline{\aL},\overline{\aM},\overline{\aE},\overline{\aS}\}$ is equivalent to the original one.
\section{Mathematical modelling of complex systems using GENERIC}
\label{sec: modelling}
As explained in Section \ref{sec: GENERIC}, the GENERIC framework provides a systematic method to derive thermodynamically consistent evolution equations. The modelling procedure of complex systems using GENERIC can be summarized in the following procedure consisting of three steps.
\begin{enumerate}[Step 1.]
\item Choose state variable $\az\in \aZ$;
\item Choose (construct) GENERIC building blocks that include
\begin{itemize}
\item two functionals $\aE$ and $\aS$
\item two operators $\aL$ and $\aM$
\end{itemize}
such that the GENERIC conditions (cf. Section \ref{sec: GENERIC}) are fulfilled;
\item Derive the equation.
\end{enumerate}
We emphasize again that the quintuple $\{\aZ,\aE,\aS,\aL,\aM\}$ completely determine the model and that the thermodynamics laws are automatically justified in the GENERIC framework. 

In this paper, we apply this procedure to cast the existing damped Timoshenko and damped Bresse systems into the GENERIC and to derive new models.

All the existing damped Timoshenko and damped Bresse models share a common feature: they are described by wave equations coupled with damping mechanisms, either by frictional or heat conduction. It is well-known that, without the damping effect, the energy of a wave equation is conserved. In other words, a wave equation is conservative. However, when the damping effect is present, the energy is no longer conserved.  This implies that the damped Timoshenko and damped Bresse systems exhibit both conservative and dissipative effects represented in the wave equations and the damping mechanisms respectively. It is the reason why GENERIC would be a natural framework to work with. To place them in the GENERIC setting, we need to specify the building blocks and verify the conditions imposed on them, cf. Section \ref{sec: GENERIC}. As also mentioned in Section~\ref{sec: GENERIC}, one might obtain the same GENERIC system from different building blocks provided that they are related by a co-ordinate transformation~\eqref{eq: transform}. Therefore, one can mathematically simplify the construction of the building blocks by choosing the state space in such a way that the entropy is multiple of one of its components. This technique has been used in~\cite{DPZ13a} for the Vlasov-Fokker-Planck equation and will be used throughout the present paper. 

We are now ready to introduce a common procedure to place the existing damped Timoshenko and damped Bresse systems in the GENERIC framework. This procedure is a slightly modification of the one described above where the first step is split into Step 1a and Step 1b and the second step is decomposed to Step 2a and Step 2b.
\begin{enumerate}
\item [Step 1a.] Re-write the system as a system of parabolic partial differential equations;
\item [Step 1b.] Introduce a new variable $e$, depending only on time, to capture the lost of the energy of the original system. The time derivative of $e$ is simply determined by the negative of that of the energy of the original system. 
\item [Step 2a.] Construct the building blocks of the GENERIC: the GENERIC-energy $\aE$ is defined to be the summation of the original energy and $e$; the GENERIC-entropy $\aS $ is a multiple of $e$; the Poisson operator $\aL$ is easily deduced from the wave equations, the anti-symmetry property and and the degeneracy  condition $\aL(\az)\frac{\delta \aS}{\delta \az}(\z)=0$; and the dissipative operator $\aM$ is built from the damping effect, the symmetry property and the degeneracy condition $\aM(\az)\frac{\delta \aE}{\delta \az}(\z)=0$;
\item [Step 2b.] Verify the remaining conditions;
\item [Step 3.] Derive the equation.
\end{enumerate}

\section{GENERIC formulation of damped Timoshenko systems}
\label{sec: Timoshenko}
In this section, we place damped Timoshenko systems of various types in the GENERIC framework. We perform the procedure described in Section \ref{sec: modelling} in details for two systems: the Timoshenko system with dual frictional damping and the Timoshenko system damped by heat conduction of type I. 
We also show how to derive a new GENERIC model. Some details of computation will be given in Appendix A, and the Timoshenko system damped by heat conduction of type II and type III will be presented in Appendix B. 
\subsection{The Timoshenko system}
The Timoshenko system, which describes the transverse vibrations of a beam, is a set of two coupled wave equations of the form
\begin{equation}
\label{eq: Timoshenko}
\begin{aligned}
\varphi_{tt}&=k(\varphi_x+\psi)_x,
\\ \psi_{tt}&=b\psi_{xx}-k(\varphi_x+\psi).
\end{aligned}
\end{equation}
Here, $t$ is the time variable and $x$ is the space coordinate along the beam. The function $\varphi$ is the transverse displacement and $\psi$ is the rotation angle of a filament of the beam. Throughout this paper, subscripts denote derivatives of the functions with respect to the corresponding variables; for instance, $\varphi_x$ is the first order derivative of $\varphi$ with respect to $x$. Finally, $k,b$ are positive constants.

The energy of the beam is
\[
E(t)=\int_{\Omega}\left(\frac{1}{2}\varphi_t^2+\frac{1}{2}\psi_t^2+\frac{k}{2}(\varphi_x+\psi)^2+\frac{b}{2}\psi_x^2\right)\,dx.
\]
Solutions of~\eqref{eq: Timoshenko} do not decay and the system's energy remains constant at all times. This can be seen easily by computing the derivative of $E(t)$ with respect to time
\begin{align*}
\frac{d}{dt}E(t)&=\int_{\Omega} [\varphi_t \varphi_{tt}+\psi_t\psi_{tt}+k(\varphi_x+\psi)(\varphi_{xt}+\psi_t)+b\psi_x\psi_{xt}]\,dx
\\&=\int_{\Omega} [k\varphi_t(\varphi_x+\psi)_x +\psi_t(b\psi_{xx}-k(\varphi_x+\psi))+k(\varphi_x+\psi)(\varphi_{xt}+\psi_t)+b\psi_x\psi_{xt}]\,dx
\\&=0.
\end{align*}

There is a large amount of research in the literature devoted to find the damping necessary to add to the system~\eqref{eq: Timoshenko} in order to stabilize its solutions~\cite{RR02,RFSC05,HK13,SR09,RR08,FRM14}. Two damping mechanisms have been mainly considered, namely damping by frictions and by heat conduction. The latter can be accomplished in various ways. By adding damping effects, we get damped Timoshenko systems, and the energy above is no longer preserved. The main results of the mentioned papers are that damped Timoshenko systems are exponentially stable.

In the next sections, we will show that damped Timoshenko systems, either by frictions or by heat conduction, can be cast into the GENERIC framework. 
\subsection{The Timoshenko system with dual frictional damping}
\label{sec: Timoshenko fric}
We start with the Timoshenko system~\eqref{eq: Timoshenko} with two frictional terms added, i.e., the following system 
\begin{equation}
\label{eq: double wave 1}
\begin{cases}
\varphi_{tt}=k(\varphi_x+\psi)_x-\delta_1\varphi_t,\\
\psi_{tt}=b\psi_{xx}-k(\varphi_x+\psi)-\delta_2\psi_t.
\end{cases}
\end{equation} 
The two terms $\delta_1 \varphi_t$ and $\delta_2\psi_t$, where $\delta_1,\delta_2\geq 0$, represent the frictions. This system has been studied, e.g., in~\cite{RFSC05}, where the authors prove that it is exponential stable. Set $p=\varphi_t, q=\psi_t$, then Eq.~\eqref{eq: double wave 1} can be re-written as a system of parabolic differential equations.
\begin{equation}
\label{eq: couple wave 2}
\begin{cases}
\varphi_t=p,\\
\psi_t=q,\\
p_t=-\delta_1 p+k(\varphi_x+\psi)_x,\\
q_t=-\delta_2 q-k(\varphi_x+\psi)+b\psi_{xx}.
\end{cases}
\end{equation}
We introduce an auxiliary variable $e$, \textit{depending only on $t$}, such that 
\begin{equation*}
\frac{d}{dt}e(t)+\frac{d}{dt}\int_{\Omega}\left(\frac{1}{2}p^2+\frac{k}{2}(\varphi_x+\psi)^2+\frac{1}{2}q^2+\frac{b}{2}\psi_x^2\right)\,dx=0,
\end{equation*}
which results in,
\begin{equation}
\label{eq: aux wave}
e_t=\int_{\Omega}(\delta_1p^2+\delta_2q^2)\,dx.
\end{equation}
We give more detailed discussion on the introduction of $e$ in Remark~\ref{rem: aux} below. We now show that the system~\eqref{eq: couple wave 2}-\eqref{eq: aux wave} can be cast into the GENERIC framework. The building blocks are given by
\begin{align*}
&\az=(\varphi\quad \psi\quad p\quad q \quad e)^T,\quad\aE(\az)=e+\int_\Omega\left(\frac{1}{2}p^2+\frac{k}{2}(\varphi_x+\psi)^2+\frac{1}{2}q^2+\frac{b}{2}\psi_x^2\right)dx,\quad \aS(\az)=\alpha\,e,
\\&\aL(\az)=\begin{pmatrix}
0&0&1&0&0\\
0&0&0&1&0\\
-1&0&0&0&0\\
0&-1&0&0&0\\
0&0&0&0&0
\end{pmatrix},\quad \aM(\az)=\frac{1}{\alpha}\begin{pmatrix}
0&0&0&0&0\\
0&0&0&0&0\\
0&0&\delta_1&0&-\delta_1 p\\
0&0&0&\delta_2&-\delta_2 q\\
0&0&-\delta_1 \int_{\Omega}p\cdot\square\,dx&-\delta_2 \int_{\Omega}q\cdot\square \,dx& \int_{\Omega}(\delta_1 p^2+\delta_2 q^2)\,dx
\end{pmatrix}.
\end{align*}
In the above expressions, $\alpha\in \mathbb{R}\setminus\{0\}$ is some scaled parameter. The squares $\square$ in $\aM(\az)$ represent the arguments that $\aM(\az)$ acts on. 

We now derive the GENERIC equation associated to these building blocks by computing
\begin{align*}
&\frac{\delta\aE(\az)}{\delta\az} =\begin{pmatrix}
-k(\varphi_x+\psi)_x\\
-b\psi_{xx}+k(\varphi_x+\psi)\\
p\\
q\\
1
\end{pmatrix}, \quad \frac{\delta\aS(\az)}{\delta\az} =\begin{pmatrix}
0\\
0\\
0\\
0\\
\alpha
\end{pmatrix},
\\&\aL(\az)\frac{\delta\aE(\az)}{\delta\az}=\begin{pmatrix}
p\\
q\\
k(\varphi_x+\psi)_x\\
b\psi_{xx}-k(\varphi_x+\psi)\\
0
\end{pmatrix},\quad \aM(\az)\frac{\delta\aS(\az)}{\delta\az}=\begin{pmatrix}
0\\
0\\
-\delta_1 p\\
-\delta_2 q\\
\int_{\Omega}(\delta_1 p^2+\delta_2 q^2)\,dx
\end{pmatrix}.
\end{align*}
It is now clear that the GENERIC equation
\begin{equation*}
\az_t=\aL(\az)\frac{\delta\aE(\az)}{\delta\az}+\aM(\az)\frac{\delta\aS(\az)}{\delta\az}, 
\end{equation*}
is equivalent to the system~\eqref{eq: couple wave 2}-\eqref{eq: aux wave}. Note how the GENERIC structure reveals/clarifies the conservative and dissipative parts. The verification that the building blocks $\{\aZ,\aL,\aM,\aE,\aS\}$ above satisfy the conditions of GENERIC is given in the Appendix A.
\begin{remark}
\label{rem: aux}
We emphasize that the system~\eqref{eq: couple wave 2}-\eqref{eq: aux wave} is coupled only in one direction: the newly added equation for $e$ is slaved to the original damped Timoshenko system. In other words, the introduction of $e$ does not change the original system. One can think of it as a purely mathematical technique. Alternatively, from a physical point of view, one can also interpret this as embedding the original system into a bigger reservoir/environment and $e$ captures the exchange of the energy between them.
\end{remark}
\subsection{The Timoshenko system damped by heat conduction of type I}
\label{sec: Timoshenko heat I}
In this section, we consider the Timoshenko system~\eqref{eq: Timoshenko} damped by heat conduction of the form (type I),
\begin{equation}
\label{eq: Timoshenko-Fourier}
\begin{cases}
\varphi_{tt}=k(\varphi_x+\psi)_x,\\
\psi_{tt}=b\psi_{xx}-k(\varphi_x+\psi)-\gamma\theta_x,\\
\theta_t=\kappa \theta_{xx}-\gamma \psi_{tx}.
\end{cases}
\end{equation}

In this model, $\theta$ is the temperature. It is coupled to the Timoshensko system via the term $-\gamma \theta_x$ in the equation for $\psi$. The system above, which is often known as the Timoshenko-Fourier system since the heat conduction is described by the classical Fourier law, has been studied extensively in the literature, see for instance~\cite{RR02,HK13,SR09}.
\\
Set $p=\varphi_t,q=\psi_t$, then Eq.~\eqref{eq: Timoshenko-Fourier} can be re-written as a system of parabolic differential equations
\begin{equation}
\label{eq: Timosenko-Fourier 1}
\begin{cases}
\varphi_t=p,
\psi_t=q,\\
p_t=k(\varphi_x+\psi)_x,\\
q_t=-k(\varphi_x+\psi)+b\psi_{xx}-\gamma \theta_x,\\
\theta_t=\kappa\theta_{xx}-\gamma q_x.
\end{cases}
\end{equation}

Similarly as in the previous section, we introduce an auxiliary variable $e$, depending only on $t$, such that
\begin{equation*}
\frac{d}{dt}e+\frac{d}{dt}\int_{\Omega}\left(\frac{1}{2}p^2+\frac{1}{2}q^2+\frac{k}{2}(\varphi_x+\psi)^2+\frac{b}{2}\psi_x^2+\frac{1}{2}\theta^2\right)\,dx=0,
\end{equation*}
which gives rise to
\begin{equation}
\label{eq: aux T-F}
e_t=\kappa\int_{\Omega}\theta_x^2\,dx.
\end{equation}
We now show that the coupled systems~\eqref{eq: Timosenko-Fourier 1}-\eqref{eq: aux T-F} can be cast into the GENERIC framework. The building blocks are as follows
\begin{align*}
&\az=(\varphi\quad \psi\quad p\quad q \quad \theta \quad e)^T,\quad \aE(\az)=\int_\Omega\left(\frac{1}{2}p^2+\frac{k}{2}(\varphi_x+\psi)^2+\frac{1}{2}q^2+\frac{b}{2}\psi_x^2+\frac{1}{2}\theta^2\right)dx +e,\quad \aS(\az)=\alpha\,e,
\\& \aL(\az)=\begin{pmatrix}
0&0&1&0&0&0\\
0&0&0&1&0&0\\
-1&0&0&0&0&0\\
0&-1&0&0&-\gamma \partial_x&0\\
0&0&0&-\gamma\partial_x&0&0\\
0&0&0&0&0&0
\end{pmatrix},
\quad \aM(\az)=\frac{1}{\alpha}\begin{pmatrix}
0&0&0&0&0&0\\
0&0&0&0&0&0\\
0&0&0&0&0&0\\
0&0&0&0&0&0\\
0&0&0&0&-\kappa\partial_{xx}&\kappa\theta_{xx}\\
0&0&0&0&-\kappa\int_{\Omega}\theta_x\cdot\partial_x\square\,dx&\kappa\int_{\Omega}\theta_x^2\,dx
\end{pmatrix},
\\&
\end{align*}
Having the building blocks, we can derive the GENERIC equation associated to them. A direct calculation gives
\begin{align*}
&\frac{\delta \aE(\az)}{\delta\az}=\begin{pmatrix}
-k(\varphi_x+\psi)_x\\
-b\psi_{xx}+k(\varphi_x+\psi)\\
p\\
q\\
\theta\\
1
\end{pmatrix}, \quad\frac{\delta \aS(\az)}{\delta\az}=\begin{pmatrix}
0\\
0\\
0\\
0\\
0\\
\alpha
\end{pmatrix},
\\& \aL(\az)\frac{\delta \aE(\az)}{\delta\az}=\begin{pmatrix}
p\\
q\\
k(\varphi_x+\psi)_x\\
b\psi_{xx}-k(\varphi_x+\psi)-\gamma \theta_x\\
-\gamma q_x\\
0
\end{pmatrix},\quad \aM(\az)\frac{\delta \aS(\az)}{\delta\az}=\begin{pmatrix}
0\\
0\\
0\\
0\\
\kappa\theta_{xx}\\
\kappa\int_{\Omega}\theta_x^2\, dx
\end{pmatrix}.
\end{align*}
It follows that the GENERIC equation
\begin{equation*}
\az_t=\aL(\az)\frac{\delta \aE(\az)}{\delta\az}+\aM(\az)\frac{\delta \aS(\az)}{\delta\az}, 
\end{equation*}
is the same as the coupled systems~\eqref{eq: Timosenko-Fourier 1}-\eqref{eq: aux T-F}. The verification that $\{\aL,\aM,\aE,\aS\}$ satisfy the conditions of GENERIC is presented in the Appendix A.

In Appendix B, we list two more models: the Timoshenko system damped by heat conduction of type II and type III.
\subsection{A new model}
As we have shown, in order to place each damped Timoshenko systems in the previous sections in the GENERIC framework, we need to introduce an extra auxiliary variable $e$. Motivated by~\cite{Mielke11}, we introduce the following system, which is GENERIC on its own, i.e., it is not necessary to complement with an auxiliary variable. This example shows how GENERIC can be used to build up a new model that is thermodynamically consistent.
\begin{equation}
\label{eq: Timosenko type 4}
\begin{cases}
\varphi_{tt}=k(\varphi_x+\psi)_x,
\\ \psi_{tt}=b\psi_{xx}-k(\varphi_x+\psi)+\gamma\theta_x,
\\\theta_t=\delta \theta_{xx}+\gamma \theta \psi_{tx}.
\end{cases}
\end{equation}

The difference between this model and the previous one lies in the second term of the equation for the temperature. The coupling to the mechanical part (i.e., the Timoshenko system) is now non-linear. Set $p=\varphi_t, q=\psi_t$, then Eq.~\eqref{eq: Timosenko type 4} can be re-written as
\begin{equation}
\label{eq: Timosenko 4-2}
\begin{cases}
\varphi_t=p,
\\\psi_t=q,
\\p_t=k(\varphi_x+\psi)_x,
\\q_t=-k(\varphi_x+\psi)+b\psi_{xx}+\gamma \theta_x,
\\\theta_t=\delta\theta_{xx}+\gamma \theta q_x.
\end{cases}
\end{equation}

We now show that this system is a GENERIC equation. The building blocks are defined as follows
\begin{align*}
&\az=(\varphi\quad\psi\quad p\quad q\quad\theta)^T
\\&\aE(\az)=\int_\Omega\left(\frac{1}{2}p^2+\frac{k}{2}(\varphi_x+\psi)^2+\frac{1}{2}q^2+\frac{b}{2}\psi_x^2+\theta\right)dx,\quad \aS(\az)=\int_\Omega\log \theta\,dx,
\\& \aL(\az)=\begin{pmatrix}
0&0&1&0&0\\
0&0&0&1&0\\
-1&0&0&0&0\\
0&-1&0&0&\gamma (\theta\square)_x\\
0&0&0&\gamma\theta\cdot\square_x&0
\end{pmatrix},\quad \aM(\az)=\begin{pmatrix}
0&0&0&0&0\\
0&0&0&0&0\\
0&0&0&0&0\\
0&0&0&0&0\\
0&0&0&0&-\delta(\theta\square_x)_x\end{pmatrix}.
\end{align*}
We have
\begin{align*}
&\frac{\delta \aE(\az)}{\delta\az})=\begin{pmatrix}
-k(\varphi_x+\psi)_x\\
-b\psi_{xx}+k(\varphi_x+\psi)\\
p\\
q\\
1
\end{pmatrix}, \quad \frac{\delta \aS(\az)}{\delta\az}=\begin{pmatrix}
0\\
0\\
0\\
0\\
\frac{1}{\theta}
\end{pmatrix},
\\&\aL(\az)\frac{\delta \aE(\az)}{\delta\az})=\begin{pmatrix}
p\\
q\\
k(\varphi_x+\psi)_x\\
b\psi_{xx}-k(\varphi_x+\psi)+\gamma \theta_x\\
\gamma \theta q_x
\end{pmatrix},\quad \aM(\az)\frac{\delta \aS(\az)}{\delta\az})=\begin{pmatrix}
0\\
0\\
0\\
0\\
\delta \theta_{xx}
\end{pmatrix}.
\end{align*}
Substituing the above computation to the GENERIC equation\begin{equation}
\az_t=\aL(\az)\frac{\delta \aE(\az)}{\delta\az})+\aM(\az)\frac{\delta \aS(\az)}{\delta\az}), 
\end{equation}
yields the system~\eqref{eq: Timosenko 4-2}. It can be verified analogously as in the Appendix A that $\{\aL,\aM,\aE,\aS\}$ satisfy the conditions of the GENERIC.

We now move to the second class of systems.
\section{Damped Bresse systems}
\label{sec: Bresse}
In this section, we cast damped Bresse systems into the GENERIC framework. We recall the undamped Bresse system first.
\subsection{The Bresse system}
The Bresse system, which is also known as the circular arch problem, is given by three coupled wave type equations as follows
\begin{equation}
\label{eq: orignial Bresse eqn}
\begin{cases}
\varphi_{tt}=k(\varphi_x+\psi+l \phi)_x+k_0l(\phi_x-l\varphi),\\
\psi_{tt}=b\psi_{xx}-k(\varphi_x+\psi+l\phi),\\
\phi_{tt}=k_0(\phi_x-l\varphi)_x-kl(\varphi_x+\psi+l\phi).
\end{cases}
\end{equation}
In this equation, $t$ and $x$ are respectively the time and space variables. The functions $\phi,\varphi$ and $\psi$ represent the longitudinal, vertical, and shear angle displacements of elastic materials such as flexible beams. $k,l,k_0$ and $b$ are positive constants. More information about mathematical modeling of the Bresse system can be found in~\cite{LLS93}. Eq.~\eqref{eq: orignial Bresse eqn} is conservative in the sense that it preserves the energy
\begin{equation*}
E(t)=\int_{\Omega} \left(\frac{1}{2}\varphi_t^2+\frac{1}{2}\psi_t^2+\frac{1}{2}\phi_t^2+\frac{k}{2}(\varphi_x+\psi+l \phi)^2+\frac{b}{2}\psi_x^2+\frac{k_0}{2}(\phi_x-l\varphi)^2\right)\,dx.
\end{equation*}
Like in the Timoshenko system~\eqref{eq: Timoshenko}, Eq. \eqref{eq: orignial Bresse eqn} is not stable and much research have been done to find damping effects that need to be included to stabilize~\eqref{eq: orignial Bresse eqn}. Again, two mechanisms are taken into account: frictional damping and heat conduction. 

In the next sections, we show that damped Bresse systems of various types can be placed in the GENERIC framework.

\subsection{The Bresse system with frictional damping}
In this section, we consider the Bresse system with frictional dissipation,
\begin{equation}
\label{eq: Bresse frictional}
\begin{cases}
\varphi_{tt}=k(\varphi_x+\psi+l \phi)_x+k_0l(\phi_x-l\varphi)-\gamma_1 \varphi_t,
\\\psi_{tt}=b\psi_{xx}-k(\varphi_x+\psi+l\phi)-\gamma_2\psi_t,
\\\phi_{tt}=k_0(\phi_x-l\varphi)_x-kl(\varphi_x+\psi+l\phi)-\gamma_3 \phi_t.
\end{cases}
\end{equation}
This system is \eqref{eq: orignial Bresse eqn} with three frictional terms added $\gamma_1 \varphi_t,\gamma_2\psi_t,\gamma_3 \phi_t$, where $\gamma_1,\gamma_2$ and $\gamma_3$ are non-negative constants. Stability property and rate of decay of solutions of this system have been studied in, e.g., \cite{SA10,BRJ11,FM12,AFSM14}.

We now cast this system into the GENERIC framework. Since the procedure is similar as in Section \ref{sec: Timoshenko}, we only list the computations here.
\begin{itemize}
\item Re-write the system by introducing new variables $p,q$ and $w$
\begin{equation}
\label{eq: Bresse fric 2}
\begin{cases}
\varphi_t=p,
\\\psi_t=q,
\\\phi_t=w,
\\ p_t=k(\varphi_x+\psi+l \phi)_x+k_0l(\phi_x-l\varphi)-\gamma_1 p,
\\q_{t}=b\psi_{xx}-k(\varphi_x+\psi+l\phi)-\gamma_2 q,
\\w_t=k_0(\phi_x-l\varphi)_x-kl(\varphi_x+\psi+l\phi)-\gamma_3 w.
\end{cases}
\end{equation}
\item Equation for $e$
\begin{equation}
\label{eqn for e Bresse fric}
e_t=-\frac{d}{dt}\int \left(\frac{1}{2}p^2+\frac{1}{2}q^2+\frac{1}{2}w^2+\frac{k}{2}(\varphi_x+\psi+l \phi)^2+\frac{b}{2}\psi_x^2+\frac{k_0}{2}(\phi_x-l\varphi)^2\right)\,dx=\int_{\Omega}(\gamma_1 p^2+\gamma_2 q^2+\gamma_3 w^2)\,dx.
\end{equation}
\item GENERIC building blocks
\begin{align*}
&\az=\begin{pmatrix}\varphi\quad\psi\quad\phi\quad p\quad q\quad w\quad e
\end{pmatrix}^T, 
\\& \aE(\az)= e+\int \left(\frac{1}{2}p^2+\frac{1}{2}q^2+\frac{1}{2}w^2+\frac{k}{2}(\varphi_x+\psi+l \phi)^2+\frac{b}{2}\psi_x^2+\frac{k_0}{2}(\phi_x-l\varphi)^2\right)\,dx,\quad \aS(\az)=\alpha\,e,
\\& \aL(\az)=\begin{pmatrix}
0&0&0&1&0&0&0\\
0&0&0&0&1&0&0\\
0&0&0&0&0&1&0\\
-1&0&0&0&0&0&0\\
0&-1&0&0&0&0&0\\
0&0&-1&0&0&0&0\\
0&0&0&0&0&0&0
\end{pmatrix},
\\& \aM(\az)=\frac{1}{\alpha}\begin{pmatrix}
0&0&0&0&0&0&0\\
0&0&0&0&0&0&0\\
0&0&0&0&0&0&0\\
0&0&0&\gamma_1&0&0&-\gamma_1 p\\
0&0&0&0&\gamma_2&0&-\gamma_2 q\\
0&0&0&0&0&\gamma_3&-\gamma_3 w\\
0&0&0&-\gamma_1 \int_{\Omega}p\cdot\square\,dx&-\gamma_2 \int_{\Omega}q\cdot\square \,dx&-\gamma_3 \int_{\Omega}w\cdot\square \,dx& \int_{\Omega}(\gamma_1 p^2+\gamma_2 q^2+\gamma_3 w^2)\,dx
\end{pmatrix}.
\end{align*}
\item A direct calculations gives
\begin{align*}
&\frac{\delta \aE(\az)}{\delta\az}=\begin{pmatrix}
-k(\varphi_x+\psi+l \phi)_x-k_0l(\phi_x-l\varphi)\\
-b\psi_{xx}+k(\varphi_x+\psi+l \phi)\\
-k_0(\phi_x-l\varphi)_x+kl(\varphi_x+\psi+l \phi)\\
p\\
q\\
w\\
1
\end{pmatrix},\quad\frac{\delta \aS(\az)}{\delta\az}=\begin{pmatrix}
0\\
0\\
0\\
0\\
0\\
0\\
\alpha
\end{pmatrix},
\\&\aL(\az)\frac{\delta \aE(\az)}{\delta\az}=\begin{pmatrix}
p\\
q\\
w\\
k(\varphi_x+\psi+l \phi)_x+k_0l(\phi_x-l\varphi)\\
b\psi_{xx}-k(\varphi_x+\psi+l \phi)\\
k_0(\phi_x-l\varphi)_x-kl(\varphi_x+\psi+l \phi)\\
0
\end{pmatrix},\quad \aM(\az)\frac{\delta \aS(\az)}{\delta\az}=\begin{pmatrix}
0\\
0\\
0\\
-\gamma_1 p\\
-\gamma_2 q\\
-\gamma_3 w\\
\int_{\Omega}(\gamma_1 p^2+\gamma_2 q^2+\gamma_3 w^2)\,dx
\end{pmatrix}.
\end{align*}
\end{itemize}
Then the GENERIC
\begin{equation*}
\az_t=\aL(\az)\frac{\delta \aE(\az)}{\delta\az}+\aM(\az)\frac{\delta \aS(\az)}{\delta\az}, 
\end{equation*}
is the same as the system~\eqref{eq: Bresse fric 2}-\eqref{eqn for e Bresse fric}.
\subsection{The Bresse system damped by heat conduction: type I}
In this section, we consider the Bresse system coupled to a heat conduction of the form (type I)
\begin{equation}
\label{eq: Bresse heat I 1}
\begin{cases}
\varphi_{tt}=k(\varphi_x+\psi+l \phi)_x+k_0l(\phi_x-l\varphi),\\
\psi_{tt}=b\psi_{xx}-k(\varphi_x+\psi+l\phi)-\gamma \theta_x,\\
\phi_{tt}=k_0(\phi_x-l\varphi)_x-kl(\varphi_x+\psi+l\phi),\\
\theta_t=\kappa\theta_{xx}-\gamma\psi_{xt}.
\end{cases}
\end{equation}
This system is \eqref{eq: orignial Bresse eqn} coupled to the heat conduction given by the last equation. Stability property and rate of decay of solutions of this system have been studied in, e.g., \cite{FR10,SB14}. 

We now show the steps to place this system in the GENERIC setting.
\begin{itemize}
\item Re-write the system
\begin{equation}
\label{eq: Bresse heat I 2}
\begin{cases}
\varphi_t=p,
\\\psi_t=q,
\\\phi_t=w,
\\ p_t=k(\varphi_x+\psi+l \phi)_x+k_0l(\phi_x-l\varphi),
\\q_{t}=b\psi_{xx}-k(\varphi_x+\psi+l\phi)-\gamma \theta_x,
\\w_t=k_0(\phi_x-l\varphi)_x-kl(\varphi_x+\psi+l\phi),
\\\theta_t=\kappa \theta_{xx}-\gamma q_x.
\end{cases}
\end{equation}
\item Equation for $e$
\begin{equation}
\label{eq: eq eB2}
e_t=-\frac{d}{dt}\int \left(\frac{1}{2}p^2+\frac{1}{2}q^2+\frac{1}{2}w^2+\frac{k}{2}(\varphi_x+\psi+l \phi)^2+\frac{b}{2}\psi_x^2+\frac{k_0}{2}(\phi_x-l\varphi)^2+\frac{1}{2}\theta^2\right)\,dx=\kappa\int_{\Omega}\theta_x^2\,dx.
\end{equation}
\item GENERIC building blocks
\begin{align*}
&\az=\begin{pmatrix} \varphi\quad\psi\quad \phi\quad p\quad q\quad w\quad \theta\quad e \end{pmatrix}^T, 
\\& \aE(\az)= e+\int \left(\frac{1}{2}p^2+\frac{1}{2}q^2+\frac{1}{2}w^2+\frac{k}{2}(\varphi_x+\psi+l \phi)^2+\frac{b}{2}\psi_x^2+\frac{k_0}{2}(\phi_x-l\varphi)^2+\frac{1}{2}\theta^2\right)\,dx,\quad \aS(\az)=\alpha\,e,
\\& \aL(\az)=\begin{pmatrix}
0&0&0&1&0&0&0&0\\
0&0&0&0&1&0&0&0\\
0&0&0&0&0&1&0&0\\
-1&0&0&0&0&0&0&0\\
0&-1&0&0&0&0&-\gamma \partial_x&0\\
0&0&-1&0&0&0&0&0\\
0&0&0&0&-\gamma \partial_x&0&0&0\\
0&0&0&0&0&0&0&0
\end{pmatrix},
\\& \aM(\az)=\frac{1}{\alpha}\begin{pmatrix}
0&0&0&0&0&0&0\\
0&0&0&0&0&0&0\\
0&0&0&0&0&0&0\\
0&0&0&0&0&0&0\\
0&0&0&0&0&0&0\\
0&0&0&0&0&0&0\\
0&0&0&0&0&-\kappa\partial_{xx}&\kappa\theta_{xx}\\
0&0&0&0&0&-\kappa\int_{\Omega}\theta_x\cdot\partial_x\square\,dx&\kappa\int_{\Omega}\theta_x^2\,dx
\end{pmatrix}.
\end{align*}
\item A direct calculation gives 
\begin{align*}
&\frac{\delta \aE(\az)}{\delta\az}=\begin{pmatrix}
-k(\varphi_x+\psi+l \phi)_x-k_0l(\phi_x-l\varphi)\\
-b\psi_{xx}+k(\varphi_x+\psi+l \phi)\\
-k_0(\phi_x-l\varphi)_x+kl(\varphi_x+\psi+l \phi)\\
p\\
q\\
w\\
\theta\\
1
\end{pmatrix},\quad\frac{\delta \aS(\az)}{\delta\az}=\begin{pmatrix}
0\\
0\\
0\\
0\\
0\\
0\\
0\\
\alpha
\end{pmatrix},
\\&\aL(\az)\frac{\delta \aE(\az)}{\delta\az}=\begin{pmatrix}
p\\
q\\
w\\
k(\varphi_x+\psi+l \phi)_x+k_0l(\phi_x-l\varphi)\\
b\psi_{xx}-k(\varphi_x+\psi+l \phi)-\gamma\theta_x\\
k_0(\phi_x-l\varphi)_x-kl(\varphi_x+\psi+l \phi)\\
-\gamma q_x\\
0
\end{pmatrix},\quad \aM(\az)\frac{\delta \aS(\az)}{\delta\az}=\begin{pmatrix}
0\\
0\\
0\\
0\\
0\\
0\\
\kappa\theta_{xx}\\
\kappa\int_{\Omega}\theta_x^2\,dx
\end{pmatrix}.
\end{align*}
Then it follows that the GENERIC
\begin{equation*}
\az_t=\aL(\az)\frac{\delta \aE(\az)}{\delta\az}+\aM(\az)\frac{\delta \aS(\az)}{\delta\az}, 
\end{equation*}
is the same as the system~\eqref{eq: Bresse heat I 2}-\eqref{eq: eq eB2}.
\end{itemize}
In Appendix B, we show one more model: the Bresse system damped by heat conduction of type II.
\section{Conclusion and discussion}
In this paper we have introduced a modelling procedure of complex systems using GENERIC. As a concrete example, this procedure allows us to unify many existing damped Timoshenko and damped Bresse systems into the GENERIC framework and derive a new model. This formulation not only provides an alternative thermodynamically consistent derivation but also reveals geometrical structures, via the GENERIC building blocks, of these systems. An important question for further research would be on mathematical analysis, such as well-posedness and asymptotic limits, of the damped Timoshenko and damped Bresse systems (and more general thermodynamical systems) using GENERIC structure. Recently it has become clear, see e.g., \cite{SandierSerfaty04,Serfaty11,AMPSV12} and very recent papers \cite{Mielke14,Mielke15}, that variational structure has important consequences for the analysis of an evolution equation. This is because variational structure can provide many good concepts of weak solution and many techniques in calculus of variations, such as Gamma convergence, can be exploited. There is a large literature on mathematical analysis of gradient flows, which is an instance of the GENERIC where the reversible effect is absent, using variational formulation, see the papers mentioned above and references therein. However, that of for GENERIC is still lacking. A variational formulation for a full GENERIC system has been proposed recently in \cite{DPZ13b}. We expect that this variational can be used for the GENERIC systems studied in this paper.
\label{sec: summary}
\section{Appendix A: Verification of the GENERIC conditions}
In this Appendix, we present the verification of the GENERIC conditions for the Timoshenko system with two frictional damping and the Timoshenko system damped by heat conduction of type I. The verification for the other models are similar and hence omitted.
\subsection{The Timoshenko system with dual frictional damping }
\label{sec: verify 1}
We now verify that $\{\aL,\aM,\aE,\aS\}$ constructed in Section~\ref{sec: Timoshenko fric} satisfy the conditions of the GENERIC. Let $\aF, \aG: \aZ\rightarrow \R$ be given.
We have
\begin{align*}
\{\aF,\aG\}_\aL&=\frac{\delta\aF(\az)}{\delta\az}\cdot \aL(\az)\,\frac{\delta\aG(\az)}{\delta\az}=\begin{pmatrix}
\frac{\delta\aF(\az)}{\delta \varphi}\\
\frac{\delta\aF(\az)}{\delta \psi}\\
\frac{\delta\aF(\az)}{\delta p}\\
\frac{\delta\aF(\az)}{\delta q}\\
\frac{\delta\aF(\az)}{\delta e}
\end{pmatrix}\cdot\begin{pmatrix}
\frac{\delta\aG(\az)}{\delta p}\\
\frac{\delta\aG(\az)}{\delta q}\\
-\frac{\delta\aG(\az)}{\delta \varphi}\\
-\frac{\delta\aG(\az)}{\delta \psi}\\
0
\end{pmatrix}
\\&=\int_{\Omega}\left[\frac{\delta\aF(\az)}{\delta \varphi}\frac{\delta\aG(\az)}{\delta p}+\frac{\delta\aF(\az)}{\delta \psi}\frac{\delta\aG(\az)}{\delta q}-\frac{\delta\aF(\az)}{\delta p}\frac{\delta\aG(\az)}{\delta \varphi}-\frac{\delta\aF(\az)}{\delta q}\frac{\delta\aG(\az)}{\delta \psi}\right]\,dx
\\&=-\int_{\Omega}\left[\frac{\delta\aG(\az)}{\delta \varphi}\frac{\delta\aF(\az)}{\delta p}+\frac{\delta\aG(\az)}{\delta \psi}\frac{\delta\aF(\az)}{\delta q}-\frac{\delta\aG(\az)}{\delta p}\frac{\delta\aF(\az)}{\delta \varphi}-\frac{\delta\aG(\az)}{\delta q}\frac{\delta\aF(\az)}{\delta \psi}\right]\,dx
=-\{\aG,\aF\}_\aL,
\end{align*}
i.e., $\aL$ is anti-symmetric. 

The verification that $\aL$ satisfies the Jacobi identity can be done by computing $\{\{\aF_1,\aF_2\}_{\aL},\aF_3\}_{\aL}+\{\{\aF_2,\aF_3\}_{\aL},\aF_1\}_{\aL}+\{\{\aF_3,\aF_1\}_{\aL},\aF_2\}_{\aL}$ directly similarly as above. The computation is lengthy and tedious, hence it is omitted. It seems that there is no general method to verify the Jacobi identity. According to~\cite[Chapter 9]{Gol01} ``there seems to be no simple way of proving Jacobi's identity for the Poisson bracket without lengthy algebra.". However, it should be mentioned that Kr\"{o}ger and H\"{u}tter  \cite{KH10} have developed a Mathematica notebook which facilitates verification of the Jacobi identity.
Next, we check conditions on $\aM$. Without loss of generality and for simplicity of notation, we set $\alpha=1$ throughout this Appendix. We have
\begin{align*}
[\aF,\aG]_\aM&=\frac{\delta\aF(\az)}{\delta\az}\cdot \aM(\az)\,\frac{\delta\aG(\az)}{\delta\az}
\\&=\begin{pmatrix}
\frac{\delta\aF(\az)}{\delta \varphi}\\
\frac{\delta\aF(\az)}{\delta \psi}\\
\frac{\delta\aF(\az)}{\delta p}\\
\frac{\delta\aF(\az)}{\delta q}\\
\frac{\delta\aF(\az)}{\delta e}
\end{pmatrix}\cdot\begin{pmatrix}
0\\
0\\
\delta_1\frac{\delta\aG(\az)}{\delta p}-\delta_1 p\frac{\delta\aG(\az)}{\delta e}\\
\delta_2\frac{\delta\aG(\az)}{\delta q}-\delta_2 q\frac{\delta\aG(\az)}{\delta e}\\
\int_{\Omega}\left[-\delta_1 p\frac{\delta\aG(\az)}{\delta p}-\delta_2 q\frac{\delta\aG(\az)}{\delta q}+\frac{\delta\aG(\az)}{\delta e}(\delta_1 p^2+\delta_2 q^2)\right]\, dx\end{pmatrix}
\\&=\int_{\Omega}\left[\delta_1\frac{\delta\aF(\az)}{\delta p}\left(\frac{\delta\aG(\az)}{\delta p}- p\frac{\delta\aG(\az)}{\delta e}\right)+\delta_2\frac{\delta\aF(\az)}{\delta q}\left(\frac{\delta\aG(\az)}{\delta q}- q\frac{\delta\aG(\az)}{\delta e}\right)\right]\,dx
\\&\qquad+\frac{\delta\aF(\az)}{\delta e}\int_{\Omega}\left[-\delta_1 p\frac{\delta\aG(\az)}{\delta p}-\delta_2 q\frac{\delta\aG(\az)}{\delta q}+\frac{\delta\aG(\az)}{\delta e}(\delta_1 p^2+\delta_2 q^2)\right]\,dx
\\&=\int_{\Omega}\left[\delta_1\frac{\delta\aG(\az)}{\delta p}\left(\frac{\delta\aF(\az)}{\delta p}- p\frac{\delta\aF(\az)}{\delta e}\right)+\delta_2\frac{\delta\aG(\az)}{\delta q}\left(\frac{\delta\aF(\az)}{\delta q}- q\frac{\delta\aF(\az)}{\delta e}\right)\right]\,dx
\\&\qquad+\frac{\delta\aG(\az)}{\delta e}\int_{\Omega}\left[-\delta_1 p\frac{\delta\aF(\az)}{\delta p}-\delta_2 q\frac{\delta\aF(\az)}{\delta q}+\frac{\delta\aF(\az)}{\delta e}(\delta_1 p^2+\delta_2 q^2)\right]\,dx
\\&=[\aG,\aF]_\aM,
\end{align*}
i.e., $\aM$ is symmetric. It also follows from the above computation that
\begin{equation*}
[\aF,\aF]_{\aM}=\int_{\Omega}\left[\delta_1\left(\frac{\delta\aF(\az)}{\delta p}-p\frac{\delta\aF(\az)}{\delta e}\right)^2+\delta_2\left(\frac{\delta\aF(\az)}{\delta q}-q\frac{\delta\aF(\az)}{\delta e}\right)^2\right]\,dx\geq 0,
\end{equation*}
i.e., $\aM$ is positive semidefinite. It remains to verify the degeneracy conditions. Indeed,
\begin{align*}
&\aM(\az)\frac{\delta\aE(\az)}{\delta\az}=\begin{pmatrix}
0&0&0&0&0\\
0&0&0&0&0\\
0&0&\delta_1&0&-\delta_1 p\\
0&0&0&\delta_2&-\delta_2 q\\
0&0&-\delta_1 \int_{\Omega}p\cdot\square\,dx&-\delta_2 \int_{\Omega}q\cdot\square \,dx& \int_{\Omega}(\delta_1 p^2+\delta_2 q^2)\,dx
\end{pmatrix}\cdot \begin{pmatrix}
-k(\varphi_x+\psi)_x\\
-b\psi_{xx}+k(\varphi_x+\psi)\\
p\\
q\\
1
\end{pmatrix}=0,
\\&\aL(\az)\frac{\delta\aS(\az)}{\delta\az}=\begin{pmatrix}
0&0&1&0&0\\
0&0&0&1&0\\
-1&0&0&0&0\\
0&-1&0&0&0\\
0&0&0&0&0
\end{pmatrix}\cdot \begin{pmatrix}
0\\
0\\
0\\
0\\
1
\end{pmatrix}=0.
\end{align*}
\subsection{The Timoshenko system damped by heat conduction of type I}
\label{sec: verify 2}
In this section, we verify that $\{\aL,\aM,\aE,\aS\}$ constructed in Section~\ref{sec: Timoshenko heat I} satisfy the conditions of the GENERIC. Let $\aF, \aG: \aZ\rightarrow \R$ be given.
We have
\begin{align*}
&\{\aF,\aG\}_\aL=\frac{\delta\aF(\az)}{\delta\az}\cdot \aL(\az)\,\frac{\delta\aG(\az)}{\delta\az}=\begin{pmatrix}
\frac{\delta\aF(\az)}{\delta \varphi}\\
\frac{\delta\aF(\az)}{\delta \psi}\\
\frac{\delta\aF(\az)}{\delta p}\\
\frac{\delta\aF(\az)}{\delta q}\\
\frac{\delta\aF(\az)}{\delta \theta}\\
\frac{\delta\aF(\az)}{\delta e}
\end{pmatrix}\cdot\begin{pmatrix}
\frac{\delta\aG(\az)}{\delta p}\\
\frac{\delta\aG(\az)}{\delta q}\\
-\frac{\delta\aG(\az)}{\delta \varphi}\\
-\frac{\delta\aG(\az)}{\delta \psi}-\gamma\frac{\partial}{\partial x}\left(\frac{\delta\aG(\az)}{\delta \theta}\right)\\
-\gamma\frac{\partial}{\partial x}\left(\frac{\delta\aG(\az)}{\delta q}\right)\\
0
\end{pmatrix}
\\ &\quad =\int_{\Omega}\left[\frac{\delta\aF(\az)}{\delta \varphi}\frac{\delta\aG(\az)}{\delta p}+\frac{\delta\aF(\az)}{\delta \psi}\frac{\delta\aG(\az)}{\delta q}-\frac{\delta\aF(\az)}{\delta p}\frac{\delta\aG(\az)}{\delta \varphi}-\frac{\delta\aF(\az)}{\delta q}\frac{\delta\aG(\az)}{\delta \psi}-\gamma \frac{\delta\aF(\az)}{\delta q}\partial_x\frac{\delta\aG(\az)}{\delta \theta}- \gamma \frac{\delta\aF(\az)}{\delta \theta}\partial_x\frac{\delta\aG(\az)}{\delta q}\right]\,dx
\\&\quad =-\int_{\Omega}\left[\frac{\delta\aG(\az)}{\delta \varphi}\frac{\delta\aF(\az)}{\delta p}+\frac{\delta\aG(\az)}{\delta \psi}\frac{\delta\aF(\az)}{\delta q}-\frac{\delta\aG(\az)}{\delta p}\frac{\delta\aF(\az)}{\delta \varphi}-\frac{\delta\aG(\az)}{\delta q}\frac{\delta\aF(\az)}{\delta \psi}-\gamma\frac{\delta\aG(\az)}{\delta q}\partial_x\frac{\delta\aF(\az)}{\delta \theta}-\gamma\frac{\delta\aG(\az)}{\delta \theta}\partial_x\frac{\delta\aF(\az)}{\delta q}\right]\,dx
\\&\quad 
=-\{\aG,\aF\}_\aL,
\end{align*}
i.e., $\aL$ is anti-symmetric. The verification that $\aL$ satisfies the Jacobi identity can be done by computing $\{\{\aF_1,\aF_2\}_{\aL},\aF_3\}_{\aL}+\{\{\aF_2,\aF_3\}_{\aL},\aF_1\}_{\aL}+\{\{\aF_3,\aF_1\}_{\aL},\aF_2\}_{\aL}$ directly similarly as above. The computation is lengthy and tedious, hence it is omitted.
\begin{align*}
[\aF,\aG]_\aM&=\frac{\delta\aF(\az)}{\delta\az}\cdot \aM(\az)\,\frac{\delta\aG(\az)}{\delta\az}
\\&=\begin{pmatrix}
\frac{\delta\aF(\az)}{\delta \varphi}\\
\frac{\delta\aF(\az)}{\delta \psi}\\
\frac{\delta\aF(\az)}{\delta p}\\
\frac{\delta\aF(\az)}{\delta q}\\
\frac{\delta\aF(\az)}{\delta \theta}\\
\frac{\delta\aF(\az)}{\delta e}
\end{pmatrix}\cdot\begin{pmatrix}
0\\
0\\
0\\
0\\
-\kappa \partial_{xx}\frac{\delta\aG(\az)}{\delta \theta}+\kappa\theta_{xx}\frac{\delta\aG(\az)}{\delta e}\\
\kappa\int_{\Omega}\left[-\theta_x\partial_x\frac{\delta\aG(\az)}{\delta \theta}+\frac{\delta\aG(\az)}{\delta e}\theta_x^2\right]
\end{pmatrix}
\\&=\kappa\int_{\Omega}\left[\frac{\delta\aF(\az)}{\delta \theta}\left(-\partial_{xx}\frac{\delta\aG(\az)}{\delta \theta}+\theta_{xx}\frac{\delta\aG(\az)}{\delta e}\right)+\frac{\delta\aF(\az)}{\delta e}\left(-\theta_x\partial_x\frac{\delta\aG(\az)}{\delta \theta}+\frac{\delta\aG(\az)}{\delta e}\theta_x^2\right)\right]\,dx
\\&=\kappa\int_{\Omega}\left[\frac{\delta\aG(\az)}{\delta \theta}\left(-\partial_{xx}\frac{\delta\aF(\az)}{\delta \theta}+\theta_{xx}\frac{\delta\aF(\az)}{\delta e}\right)+\frac{\delta\aG(\az)}{\delta e}\left(-\theta_x\partial_x\frac{\delta\aF(\az)}{\delta \theta}+\frac{\delta\aF(\az)}{\delta e}\theta_x^2\right)\right]\,dx
\\&=[\aG,\aF]_\aM,
\end{align*}
i.e., $\aM$ is symmetric. It also follows from the above computation that
\begin{equation*}
[\aF,\aF]_{\aM}=\kappa\int_{\Omega}\left(\partial_x\frac{\delta\aF(\az)}{\delta \theta}-\frac{\delta\aF(\az)}{\delta e}\theta_x\right)^2\,dx\geq 0,
\end{equation*}
i.e., $\aM$ is positive semi-definite. The degeneracy condition also can be verified. Indeed,
\begin{align*}
&\aM(\az)\frac{\delta\aE(\az)}{\delta\az}=\begin{pmatrix}
0&0&0&0&0&0\\
0&0&0&0&0&0\\
0&0&0&0&0&0\\
0&0&0&0&0&0\\
0&0&0&0&-\kappa\partial_{xx}&\kappa\theta_{xx}\\
0&0&0&0&-\kappa\int_{\Omega}\theta_x\cdot\partial_x&\kappa\int_{\Omega}\theta_x^2\,dx
\end{pmatrix}\cdot \begin{pmatrix}
-k(\varphi_x+\psi)_x\\
-b\psi_{xx}+k(\varphi_x+\psi)\\
p\\
q\\
\theta\\
1
\end{pmatrix}=0,
\\&\aL(\az)\frac{\delta\aS(\az)}{\delta\az}=\begin{pmatrix}
0&0&1&0&0&0\\
0&0&0&1&0&0\\
-1&0&0&0&0&0\\
0&-1&0&0&-\gamma \partial_x&0\\
0&0&0&-\gamma\partial_x&0&0\\
0&0&0&0&0&0
\end{pmatrix}\cdot \begin{pmatrix}
0\\
0\\
0\\
0\\
0\\
1
\end{pmatrix}=0.
\end{align*}
\section{Appendix B: Other models}
\subsection{The Timoshenko system damped by heat conduction: type II}
In this section, we cast the Timoshenko system damped by heat conduction of type II into the GENERIC framework. The steps are summarized as follows.
\begin{itemize}
\begin{minipage}[t]{0.5\textwidth}
\item The original system 
\begin{equation*}
\begin{cases}
\varphi_{tt}=k(\varphi_x+\psi)_x,\\
 \psi_{tt}=b\psi_{xx}-k(\varphi_x+\psi)-\gamma\theta_x,\\
\theta_{t}=-s_{x}-\gamma\psi_{tx},\\
s_t=-\theta_x-\beta s.
\end{cases}
\end{equation*}
\end{minipage}
\begin{minipage}[t]{0.5\textwidth}
Re-write the system
\begin{equation}
\label{type II}
\begin{cases}
\varphi_t=p,\\
\psi_t=q,\\
p_t=k(\varphi_x+\psi)_x,\\
q_t=-k(\varphi_x+\psi)+b\psi_{xx}-\gamma \theta_x,\\
\theta_t=-s_x-\gamma q_x,\\
s_t=-\theta_x-\beta s.
\end{cases}
\end{equation}
\end{minipage}
Here $\theta$ and $s$ are respectively the temperature difference and the heat flux. The heat conduction in this model is described by the Cattaneo law, see for instance~\cite{RR02,HK13,SR09}.
\item Equation for $e$
\begin{equation}
\label{eq for e type II}
e_t=-\frac{d}{dt}\int_{\Omega}\left(\frac{1}{2}p^2+\frac{1}{2}q^2+\frac{k}{2}(\varphi_x+\psi)^2+\frac{b}{2}\psi_x^2+\frac{1}{2}\theta^2+\frac{1}{2}s^2\right)\,dx=\beta\int_{\Omega}s^2\,dx.
\end{equation}
\item GENERIC building blocks
\begin{align*}
&\az=(
\varphi \quad \psi\quad p\quad q\quad \theta\quad s\quad e)^T
\\&\aE(\az)=\int_\Omega\left(\frac{1}{2}p^2+\frac{k}{2}(\varphi_x+\psi)^2+\frac{1}{2}q^2+\frac{b}{2}\psi_x^2+\frac{1}{2}\theta^2+\frac{1}{2}s^2\right)dx +e,\quad \aS(\az)=\alpha\,e,
\\& \aL(\az)=\begin{pmatrix}
0&0&1&0&0&0&0\\
0&0&0&1&0&0&0\\
-1&0&0&0&0&0&0\\
0&-1&0&0&-\gamma \partial_x&0&0\\
0&0&0&-\gamma\partial_x&0&-\partial_x&0\\
0&0&0&0&-\partial_x&0&0&\\
0&0&0&0&0&0&0
\end{pmatrix},
\\&\aM(\az)=\frac{1}{\alpha}\begin{pmatrix}
0&0&0&0&0&0&0\\
0&0&0&0&0&0&0\\
0&0&0&0&0&0&0\\
0&0&0&0&0&0&0\\
0&0&0&0&0&0&0\\
0&0&0&0&0&\beta&-\beta s\\
0&0&0&0&0&-\beta\int_{\Omega}s\cdot\square\,dx&\beta\int_{\Omega}s^2\,dx,
\end{pmatrix}.
\end{align*}
\item Computation \begin{align*}
&\frac{\delta \aE(\az)}{\delta\az}=\begin{pmatrix}
-k(\varphi_x+\psi)_x\\
-b\psi_{xx}+k(\varphi_x+\psi)\\
p\\
q\\
\theta\\
s\\
1
\end{pmatrix}, \quad\frac{\delta \aS(\az)}{\delta\az}=\begin{pmatrix}
0\\
0\\
0\\
0\\
0\\
0\\
\alpha
\end{pmatrix},
\\&\aL(\az)\frac{\delta \aE(\az)}{\delta\az}=\begin{pmatrix}
p\\
q\\
k(\varphi_x+\psi)_x\\
b\psi_{xx}-k(\varphi_x+\psi)-\gamma \theta_x\\
-\gamma q_x-s_x\\
-\theta_x\\
0
\end{pmatrix},\quad \aM(\az)\frac{\delta \aS(\az)}{\delta\az}=\begin{pmatrix}
0\\
0\\
0\\
0\\
0\\
-\beta s\\
\beta\int_{\Omega}s^2\, dx
\end{pmatrix}.
\end{align*}
From the above computation, we obtain that the GENERIC equation with building $\{\az,\aL,\aM,\aE,\aS\}$ is equivalent to the system \eqref{type II}-\eqref{eq for e type II}.
\end{itemize}
\begin{remark}
The last two equations in~\eqref{type II} can be coupled to get one equation for $\theta$ as follows
\[
\theta_{tt}=\theta_{xx}-\beta \theta_t-\beta\,\gamma\psi_{tx}-\gamma \psi_{ttx}.
\]
One should compare and see the difference between the above equation and the equation for the heat conduction of the model in the next subsection.
\end{remark}

\subsection{The Timoshenko system damped by heat conduction: type III}
We now consider the Timoshenko system damped by heat conduction of the form (type III),
\begin{itemize}
\begin{minipage}[t]{0.5\textwidth}
\item The original system 
\begin{equation}
\label{eq: Timosenko type 3}
\begin{cases}
\varphi_{tt}=k(\varphi_x+\psi)_x,\\
\psi_{tt}=b\psi_{xx}-k(\varphi_x+\psi)-\beta\theta_{tx},\\
\theta_{tt}=\delta \theta_{xx}-\gamma \psi_{tx}+K\theta_{txx}.
\end{cases}
\end{equation}
\end{minipage}
\begin{minipage}[t]{0.5\textwidth}
Re-write the system
\begin{equation}
\label{eq: Timosenko type 3-2}
\begin{cases}
\varphi_t=p,\\
\psi_t=q,\\
p_t=k(\varphi_x+\psi)_x,\\
q_t=-k(\varphi_x+\psi)+b\psi_{xx}-\gamma w_x,\\
\theta_t=w,\\
w_t=\delta\theta_{xx}-\gamma q_x+Kw_{xx}.
\end{cases}
\end{equation}
\end{minipage}
This system has been investigated, e.g., in~\cite{FRM14}.
\item Equation for $e$
\begin{equation}
\label{eqn for e type III}
e_t=-\frac{d}{dt}\int_{\Omega}\left(\frac{1}{2}p^2+\frac{1}{2}q^2+\frac{1}{2}w^2+\frac{k}{2}(\varphi_x+\psi)^2+\frac{b}{2}\psi_x^2+\frac{\delta}{2}\theta_x^2\right)\,dx=K\int_{\Omega}w_x^2\,dx.
\end{equation}
\item GENERIC building blocks
\begin{align*}
&\az=(\varphi\quad\psi\quad p\quad q\quad\theta\quad w\quad e)^T
\\&\aE(\az)=\int_\Omega\left(\frac{1}{2}p^2+\frac{k}{2}(\varphi_x+\psi)^2+\frac{1}{2}q^2+\frac{b}{2}\psi_x^2+\frac{\delta}{2}\theta_x^2\right)dx +e,\quad \aS(\az)=\alpha\,e,
\\& \aL(\az)=\begin{pmatrix}
0&0&1&0&0&0&0\\
0&0&0&1&0&0&0\\
-1&0&0&0&0&0&0\\
0&-1&0&0&0&-\gamma \partial_x&0\\
0&0&0&0&0&1&0\\
0&0&0&-\gamma \partial_x&-1&0&0&\\
0&0&0&0&0&0&0
\end{pmatrix},
\\&\aM(\az)=\frac{1}{\alpha}\begin{pmatrix}
0&0&0&0&0&0&0\\
0&0&0&0&0&0&0\\
0&0&0&0&0&0&0\\
0&0&0&0&0&0&0\\
0&0&0&0&0&0&0\\
0&0&0&0&0&-K\partial_{xx}&K w_{xx}\\
0&0&0&0&0&-K\int_{\Omega}w_x\cdot\partial_x\square ~dx&K\int_{\Omega}w_x^2\,dx
\end{pmatrix},
\end{align*}
\item Computation
\begin{align*}
&\frac{\delta \aE(\az)}{\delta\az}=\begin{pmatrix}
-k(\varphi_x+\psi)_x\\
-b\psi_{xx}+k(\varphi_x+\psi)\\
p\\
q\\
-\delta\theta_{xx}\\
w\\
1
\end{pmatrix}, \quad\frac{\delta \aS(\az)}{\delta\az}=\begin{pmatrix}
0\\
0\\
0\\
0\\
0\\
0\\
\alpha
\end{pmatrix},
\\&\aL(\az)\frac{\delta \aE(\az)}{\delta\az}=\begin{pmatrix}
p\\
q\\
k(\varphi_x+\psi)_x\\
b\psi_{xx}-k(\varphi_x+\psi)-\gamma w_x\\
w\\
-\gamma q_x+\delta\theta_{xx}\\
0
\end{pmatrix},\quad \aM(\az)\frac{\delta \aS(\az)}{\delta\az}=\begin{pmatrix}
0\\
0\\
0\\
0\\
0\\
K w_{xx}\\
K\int_{\Omega}w_x^2\, dx
\end{pmatrix}.
\end{align*}
It is straightforward to verify that the GENERIC system is indeed the system \eqref{eq: Timosenko type 3-2}-\eqref{eqn for e type III}.
\end{itemize}
\subsection{The Bresse system damped by heat conduction: type II}
In this section, we investigate the Bresse system damped by two temperature equations of the following form (type II)
\begin{equation}
\label{eq: Bresse heat II 1}
\begin{cases}
\varphi_{tt}=k(\varphi_x+\psi+l \phi)_x+k_0l(\phi_x-l\varphi)-\gamma l\eta,\\
\psi_{tt}=b\varphi_{xx}-k(\varphi_x+\psi+l\phi)-\delta \theta_x,\\
\phi_{tt}=k_0(\phi_x-l\varphi)_x-kl(\varphi_x+\psi+l\phi)-\gamma \eta_x,\\
\theta_t=\kappa_1\theta_{xx}-\delta \psi_{tx},\\
\eta_t=\kappa_2\eta_{xx}-\gamma(\phi_{xt}-l\varphi_t).
\end{cases}
\end{equation}
This system has one extra equation, which is the last one, compared to~\eqref{eq: Bresse heat I 1}. Stability property and rate of decay of solutions of this system have been studied in, e.g., \cite{LR09}.
\begin{itemize}
\item Re-write the system
\begin{equation}
\label{eq: Bresse heat II 2}
\begin{cases}
\varphi_t=p,
\\\psi_t=q,
\\\phi_t=w,
\\ p_t=k(\varphi_x+\psi+l \phi)_x+k_0l(\phi_x-l\varphi)-\gamma l \eta,
\\q_{t}=b\psi_{xx}-k(\varphi_x+\psi+l\phi)-\delta \theta_x,
\\w_t=k_0(\phi_x-l\varphi)_x-kl(\varphi_x+\psi+l\phi)-\gamma\eta_x,
\\\theta_t=\kappa_1 \theta_{xx}-\delta q_x,
\\\eta_t=\kappa_2 \eta_{xx}-\gamma(w_x-l p).
\end{cases}
\end{equation}
\item Equation for $e$
\begin{equation}
\label{eq: eq eB3}
e_t=\frac{d}{dt}\int \left(\frac{1}{2}p^2+\frac{1}{2}q^2+\frac{1}{2}w^2+\frac{k}{2}(\varphi_x+\psi+l \phi)^2+\frac{b}{2}\psi_x^2+\frac{k_0}{2}(\phi_x-l\varphi)^2+\frac{1}{2}\theta^2+\frac{1}{2}\eta^2\right)\,dx=\int_{\Omega}(\kappa_1\theta_x^2+\kappa_2\eta_x^2)\,dx.
\end{equation}
\item GENERIC building block
\begin{align*}
&\az=\begin{pmatrix}\varphi\quad\psi\quad\phi\quad p\quad q\quad w\quad \theta\quad \eta\quad e\end{pmatrix}^T,
\\& \aE(\az)= e+\int \left(\frac{1}{2}p^2+\frac{1}{2}q^2+\frac{1}{2}w^2+\frac{k}{2}(\varphi_x+\psi+l \phi)^2+\frac{b}{2}\psi_x^2+\frac{k_0}{2}(\phi_x-l\varphi)^2+\frac{1}{2}\theta^2+\frac{1}{2}\eta^2\right)\,dx,\quad \aS(\az)=\alpha\,e,
\\& \aL(\az)=\begin{pmatrix}
0&0&0&1&0&0&0&0&0\\
0&0&0&0&1&0&0&0&0\\
0&0&0&0&0&1&0&0&0\\
-1&0&0&0&0&0&0&-\gamma l&0\\
0&-1&0&0&0&0&-\delta \partial_x&0&0\\
0&0&-1&0&0&0&0&-\gamma \partial_x&0\\
0&0&0&0&-\delta \partial_x&0&0&0&0\\
0&0&0&\gamma l&0&-\gamma\partial_x&0&0&0\\
0&0&0&0&0&0&0&0&0
\end{pmatrix},
\\& \aM(\az)=\frac{1}{\alpha}\begin{pmatrix}
0&0&0&0&0&0&0&0&0\\
0&0&0&0&0&0&0&0&0\\
0&0&0&0&0&0&0&0&0\\
0&0&0&0&0&0&0&0&0\\
0&0&0&0&0&0&0&0&0\\
0&0&0&0&0&0&0&0&0\\
0&0&0&0&0&0&-\kappa_1\partial_{xx}&0&\kappa_1\theta_{xx}\\
0&0&0&0&0&0&0&-\kappa_2\partial_{xx}&\kappa_2\eta_{xx}\\
0&0&0&0&0&0&-\kappa_1\int_{\Omega}\theta_x\cdot\partial_x\square\,dx&-\kappa_1\int_{\Omega}\theta_x\cdot\partial_x\square \,dx&\int_{\Omega}(\kappa_1\theta_x^2+\kappa_2\eta_x^2)\,dx
\end{pmatrix}.
\end{align*}
\end{itemize}
Next we place the coupled system \eqref{eq: Bresse heat II 2}-\eqref{eq: eq eB3} in the GENERIC framework with the building blocks,
We compute each term on the right hand side of the GENERIC equation 
\begin{align*}
&\frac{\delta \aE(\az)}{\delta\az}=\begin{pmatrix}
-k(\varphi_x+\psi+l \phi)_x-k_0l(\phi_x-l\varphi)\\
-b\psi_{xx}+k(\varphi_x+\psi+l \phi)\\
-k_0(\phi_x-l\varphi)_x+kl(\varphi_x+\psi+l \phi)\\
p\\
q\\
w\\
\theta\\
\eta\\
1
\end{pmatrix},\quad\frac{\delta \aS(\az)}{\delta\az}=\begin{pmatrix}
0\\
0\\
0\\
0\\
0\\
0\\
0\\
0\\
\alpha
\end{pmatrix},
\\&\aL(\az)\frac{\delta \aE(\az)}{\delta\az}=\begin{pmatrix}
p\\
q\\
w\\
k(\varphi_x+\psi+l \phi)_x+k_0l(\phi_x-l\varphi)-\gamma l\eta\\
b\psi_{xx}-k(\varphi_x+\psi+l \phi)-\delta\theta_x\\
k_0(\phi_x-l\varphi)_x-kl(\varphi_x+\psi+l \phi)-\gamma \eta_x\\
-\delta q_x\\
\gamma l p-\gamma w_x\\
0
\end{pmatrix},\quad \aM(\az)\frac{\delta \aS(\az)}{\delta\az}=\begin{pmatrix}
0\\
0\\
0\\
0\\
0\\
0\\
\kappa_1\theta_{xx}\\
\kappa_2\eta_{xx}\\
\int_{\Omega}(\kappa_1\theta_x^2+\kappa_2\eta_x^2)\,dx
\end{pmatrix}.
\end{align*}
Substituting these building blocks to the GENERIC equation
\begin{equation*}
\az_t=\aL(\az)\frac{\delta \aE(\az)}{\delta\az}+\aM(\az)\frac{\delta \aS(\az)}{\delta\az}, 
\end{equation*}
yields the coupled system~\eqref{eq: Bresse heat II 2}-\eqref{eq: eq eB3}.


\bibliography{bibGENERIC}

\newcommand{\etalchar}[1]{$^{#1}$}
\begin{thebibliography}{AMP{\etalchar{+}}12}

\bibitem[ADPZ11]{ADPZ11}
S.~Adams, N.~Dirr, M.~A. Peletier, and J.~Zimmer.
\newblock From a large-deviations principle to the {W}asserstein gradient flow:
  a new micro-macro passage.
\newblock {\em Comm. Math. Phys.}, 307:791--815, 2011.

\bibitem[AFSM14]{AFSM14}
M.~O. Alves, L.~H. Fatori, M.~A.~J. Silva, and R.~N. Monteiro.
\newblock Stability and optimality of decay rate for a weakly dissipative
  bresse system.
\newblock {\em Mathematical Methods in the Applied Sciences, to appear}, 2014.

\bibitem[AMP{\etalchar{+}}12]{AMPSV12}
S.~Arnrich, A.~Mielke, M.~A. Peletier, G.~Savar\'e, and M.~Veneroni.
\newblock Passing to the limit in a {W}asserstein gradient flow: From diffusion
  to reaction.
\newblock {\em Calc. Var. Partial Differential Equations}, 44(3-4):419--454,
  2012.

\bibitem[BRJ11]{BRJ11}
F.~A. Boussouira, J.~E.~M. Rivera, and D.~da S.~A. J{\'u}nior.
\newblock Stability to weak dissipative {B}resse system.
\newblock {\em J. Math. Anal. Appl.}, 374(2):481--498, 2011.

\bibitem[DLR13]{DLR13}
M.~H. Duong, V.~Laschos, and M.~Renger.
\newblock Wasserstein gradient flows from large deviations of many-particle
  limits.
\newblock {\em ESAIM Control Optim. Calc. Var.}, 19(4):1166--1188, 2013.

\bibitem[DPS13]{DPS13}
M.~H. Duong, M.~A. Peletier, and U.~Sharma.
\newblock Coarse-graining and fluctuations: Two birds with one stone.
\newblock {\em Oberwolfach Report No. 59/2013}, 2013.

\bibitem[DPZ13]{DPZ13b}
M.~H. Duong, M.~A. Peletier, and J.~Zimmer.
\newblock G{ENERIC} formalism of a {V}lasov-{F}okker-{P}lanck equation and
  connection to large-deviation principles.
\newblock {\em Nonlinearity}, 26(11):2951--2971, 2013.

\bibitem[DPZ14]{DPZ13a}
M.~H. Duong, M.~A. Peletier, and J.~Zimmer.
\newblock Conservative-dissipative approximation schemes for a generalized
  kramers equation.
\newblock {\em Mathematical Methods in the Applied Sciences},
  37(16):2517--2540, 2014.

\bibitem[Duo14]{D14b}
M.H. Duong.
\newblock {\em Large deviation and variational approaches to generalized
  gradient flows}.
\newblock PhD Thesis. Eindhoven University of Technology, 2014.

\bibitem[FI13]{FI13}
I.~Füreder and P.~Ilg.
\newblock Nonequilibrium thermodynamics of the soft glassy rheology model.
\newblock {\em Phys. Rev. E}, 88:042134, 2013.

\bibitem[FM12]{FM12}
L.~H. Fatori and R.~N. Monteiro.
\newblock The optimal decay rate for a weak dissipative {B}resse system.
\newblock {\em Appl. Math. Lett.}, 25(3):600--604, 2012.

\bibitem[FR10]{FR10}
L.~H. Fatori and J.~E.~M. Rivera.
\newblock Rates of decay to weak thermoelastic {B}resse system.
\newblock {\em IMA J. Appl. Math.}, 75(6):881--904, 2010.

\bibitem[FRM14]{FRM14}
L.~H. Fatori, J.~E.~M. Rivera, and R.~N. Monteiro.
\newblock Energy decay to {T}imoshenko's system with thermoelasticity of type
  {III}.
\newblock {\em Asymptot. Anal.}, 86(3-4):227--247, 2014.

\bibitem[G{\"O}97]{OG97}
M.~Grmela and H.~C. {\"O}ttinger.
\newblock Dynamics and thermodynamics of complex fluids. {I}. {Development} of
  a general formalism.
\newblock {\em Phys. Rev. E (3)}, 56(6):6620--6632, 1997.

\bibitem[Gol01]{Gol01}
H.~Goldstein.
\newblock {\em Classical Mechanics, 3rd edition}.
\newblock 2001.

\bibitem[HK13]{HK13}
B.~S. Houari and A.~Kasimov.
\newblock Damping by heat conduction in the {T}imoshenko system: {F}ourier and
  {C}attaneo are the same.
\newblock {\em J. Differential Equations}, 255(4):611--632, 2013.

\bibitem[HS14]{SB14}
B.~S. Houari and A.~Soufyane.
\newblock The bresse system in thermoelasticity.
\newblock {\em Mathematical Methods in the Applied Sciences, to appear}, 2014.

\bibitem[HT08a]{HutterTervoort08a}
M.~H\"{u}tter and T.~A Tervoort.
\newblock Finite anisotropic elasticity and material frame indifference from a
  nonequilibrium thermodynamics perspective.
\newblock {\em Journal of Non-Newtonian Fluid Mechanics}, 152:45 -- 52, 2008.

\bibitem[HT08b]{HutterTervoort08b}
M.~H\"{u}tter and T.~A Tervoort.
\newblock Thermodynamic considerations on non-isothermal finite anisotropic
  elasto-viscoplasticity.
\newblock {\em Journal of Non-Newtonian Fluid Mechanics}, 152:53--65, 2008.

\bibitem[KH10]{KH10}
M.~Kr{\"{o}}ger and M.~H{\"{u}}tter.
\newblock Automated symbolic calculations in nonequilibrium thermodynamics.
\newblock {\em Computer Physics Communications}, 181(12):2149 -- 2157, 2010.

\bibitem[LLS93]{LLS93}
J.~E. Lagnese, G.~Leugering, and E.~J. P.~G. Schmidt.
\newblock Modelling of dynamic networks of thin thermoelastic beams.
\newblock {\em Math. Methods Appl. Sci.}, 16(5):327--358, 1993.

\bibitem[LR09]{LR09}
Z.~Liu and B.~Rao.
\newblock Energy decay rate of the thermoelastic {B}resse system.
\newblock {\em Z. Angew. Math. Phys.}, 60(1):54--69, 2009.

\bibitem[Mie11]{Mielke11}
A.~Mielke.
\newblock Formulation of thermoelastic dissipative material behavior using
  generic.
\newblock {\em Continuum Mechanics and Thermodynamics}, 23:233--256, 2011.

\bibitem[Mie14]{Mielke14}
A.~Mielke.
\newblock On evolutionary gamma convergence for gradient systems.
\newblock {\em WIAS Preprint No. 1915}, 2014.

\bibitem[Mie15]{Mielke15}
A.~Mielke.
\newblock Variational approaches and methods for dissipative material models
  with multiple scales.
\newblock {\em WIAS Preprint No. 2084}, 2015.

\bibitem[MPR14]{MPR14}
A.~Mielke, M.A. Peletier, and D.R.M. Renger.
\newblock On the relation between gradient flows and the large-deviation
  principle, with applications to markov chains and diffusion.
\newblock {\em Potential Analysis}, 41(4):1293--1327, 2014.

\bibitem[\"O14]{Ottinger14}
H.~C. \"Ottinger.
\newblock Irreversible dynamics, onsager-casimir symmetry, and an application
  to turbulence.
\newblock {\em Phys. Rev. E}, 90:042121, 2014.

\bibitem[{\"O}G97]{OG97part2}
H.~C. {\"O}ttinger and M.~Grmela.
\newblock Dynamics and thermodynamics of complex fluids. {II}. {I}llustrations
  of a general formalism.
\newblock {\em Phys. Rev. E (3)}, 56(6):6633--6655, 1997.

\bibitem[{\"{O}}tt05]{Oettinger05}
H.~C. {\"{O}}ttinger.
\newblock {\em Beyond Equilibrium Thermodynamics}.
\newblock Wiley-Interscience, 2005.

\bibitem[RFSC05]{RFSC05}
C.~A. Raposo, J.~Ferreira, M.~L. Santos, and N.~N.~O. Castro.
\newblock Exponential stability for the {T}imoshenko system with two weak
  dampings.
\newblock {\em Appl. Math. Lett.}, 18(5):535--541, 2005.

\bibitem[RR02]{RR02}
J.~E.~M. Rivera and R.~Racke.
\newblock Mildly dissipative nonlinear {T}imoshenko systems---global existence
  and exponential stability.
\newblock {\em J. Math. Anal. Appl.}, 276(1):248--278, 2002.

\bibitem[RR08]{RR08}
J.~E.~M. Rivera and R.~Racke.
\newblock Timoshenko systems with indefinite damping.
\newblock {\em J. Math. Anal. Appl.}, 341(2):1068--1083, 2008.

\bibitem[Ser11]{Serfaty11}
S.~Serfaty.
\newblock Gamma-convergence of gradient flows on {H}ilbert and metric spaces
  and applications.
\newblock {\em Discrete Contin. Dyn. Syst.}, 31(4):1427--1451, 2011.

\bibitem[SJ10]{SA10}
M.~L. Santos and D.~da S.~A. J{\'u}nior.
\newblock Numerical exponential decay to dissipative {B}resse system.
\newblock {\em J. Appl. Math.}, pages Art. ID 848620, 17, 2010.

\bibitem[SR09]{SR09}
H.~D.~F. Sare and R.~Racke.
\newblock On the stability of damped {T}imoshenko systems: {C}attaneo versus
  {F}ourier law.
\newblock {\em Arch. Ration. Mech. Anal.}, 194(1):221--251, 2009.

\bibitem[SS04]{SandierSerfaty04}
E.~Sandier and S.~Serfaty.
\newblock Gamma-convergence of gradient flows with applications to
  {G}inzburg-{L}andau.
\newblock {\em Comm. Pure Appl. Math.}, 57(12):1627--1672, 2004.

\bibitem[Tim53]{T53}
S.~P. Timoshenko.
\newblock {\em History of Strength of Materials: With a Brief Account of the
  History of Theory of Elasticity and Theory of Structures}.
\newblock McGraw-Hill, 1953.

\end{thebibliography}
\bibliographystyle{alpha}
\end{document}